\documentclass[10pt]{elsart}
\usepackage{amssymb}
\usepackage{graphicx}
\def\tn{\textnormal}
\def\hs{\hspace}
\begin{document}
\begin{frontmatter}
\title{Random walk, cluster growth, and the morphology of urban conglomerations} 
\author{M. Pica Ciamarra, A. Coniglio}
\address{Dip.to di Scienze Fisiche, Universit\`{a} di Napoli ``Federico II'' \\
INFM-Coherentia, INFN and AMRA, Napoli, Italy}

\begin{abstract}
We propose a new model of cluster growth according to which the probability that a new unit is placed in a point at a distance $r$ from the city center is a Gaussian with mean equal to the cluster radius and variance proportional to the mean, modulated by the local density $\rho(r)$. The model is analytically solvable in $d=2$ dimensions, where the density profile varies as a complementary error function. 
The model reproduces experimental observations relative to the morphology of cities, determined via an original analysis of digital maps with a very high spatial resolution, and helps understanding the emergence of vehicular traffic.

\end{abstract}
\end{frontmatter}
\section{Introduction}
Important problems related to cluster growth processes occur in a number of different disciplines, ranging from physics to biology and transportation engineering, and several microscopic models have been proposed to describe the growth of both compact clusters, like crystals or tumors~\cite{Weeks76,Eden61}, and fractal clusters, like colloidal aggregates or snowflakes~\cite{Witten81}. All of these models are characterized by the presence of an `active' zone on the surface of the cluster where the cluster growth takes place.
Despite the simplicity which characterizes the microscopic dynamics of these cluster growth models, analytical solutions for the temporal evolution and for the spatial dependency of cluster properties, like the density profile or the width of the active zone, are difficult to obtain.
To this end one usually resorts to models for the evolution of the cluster surface, such as the Kardar-Parisi-Zhang model~\cite{Kardar86}, or to extensive numerical simulations~\cite{Plischke,Meakin86,Ossadnik93}.

These numerical simulations have suggested that, at least in the case of the Eden Model and of the DLA model, the radially averaged probability $P(r,N)dr$ that the $(N+1)$th cluster unit is deposited within a shell of width $dr$ at a distance $r$ from the center of mass of the cluster is well approximated for $r$ and $N$ large by a Gaussian distribution,
\begin{equation}
\label{eq-prob-dla}
P(r,N) = \frac{1}{(2 \pi)^{1/2} \sigma_N}\exp\left[-\frac{(r-r_N)^2}{2\sigma_N^2}\right],
\end{equation}
with mean $r_N \propto N^\nu$ and variance $\sigma \propto N^{\nu'}$; the scaling exponents 
$\nu$ and $\nu'$ are model dependent.
This growth probability distribution leads to a density profile $\rho(r,N)$ of a cluster of size $N$ in $d$ dimensions,
\begin{equation}
\label{eq-rho_dla}
\rho(r,N) = \frac{1}{S_d r^{d-1}}\int_0^N P(r,N') dN',
\end{equation}
where $S_d$ is the surface area of the $d$ dimensional unit sphere ($S_2 = 2\pi$, $S_3 = \pi$),
which can be evaluated in the limit $N \to \infty$ and $r$ large~\cite{Plischke}: $\rho(r,N=\infty) \propto r^{-d+1/\nu}$. 

Starting from Eq.~\ref{eq-prob-dla} and from some considerations about the asymptotic behavior of
$P(r,N)$ in the $r \to 0$ and $r \to \infty$ limits, in this paper we elaborate a new model for
the growth of compact clusters, 
the random walk growth model (RWG). This is based on the simple idea that the cluster radius grows as a random walker subject to a drift, which gives a growing probability distribution $P_{RWG}(r,N) \propto r^{d-1}P(r,N)$.
We solve the model in $d=2$ dimensions, showing that the density profile varies as a complementary error function. As an application of the proposed model we have studied the morphology of several European cities, which are growing clusters. Via an original analysis of digital maps with a very high resolution~\cite{maporama} we have determined the spatial dependence of their density of streets $\rho_s(r)$, which appears to be very well described by the RWG model.

\section{The Random Walk Cluster Growth Model}
In a large number of cluster growth models (DLA, Eden, Solid-on-solid, Random Deposition, $\ldots$), a cluster grows
as a new cluster unit is placed near an existing one. Therefore, in order for a cluster to grow in a given location $\vec r$, at least a cluster unit must be present near $\vec r$. In this respect, it is suprising that the growth probability of Eq.~\ref{eq-prob-dla} depends on the cluster density $\rho(r,N)$ only through $r_N$ and $\sigma_N$; instead, one would have expected the radially averaged probability of placing a cluster unit in a shell at a distance $r$ from the cluster center to be proportional to the number of cluster units which occupy the shell, i.e.  $P(r,N) \propto \rho(r,N)^{d-1} \propto r^{D_f-1}$, where $D_f=1/\nu$ is the fractal dimension of the cluster. 

We want also to point out that if $P(r,N) \propto r^{D_f-1}$, then  Eq.~\ref{eq-rho_dla} predicts the cluster density to diverge as $\rho(r) \propto r^{D_f-d}$ when $r \to 0$. We therefore expect a crossover in $P(r,N)$ which must be proportional to $r^{d-1}$ when $r \to 0$, and proportional to $r^{D_f-1}$ when $r \to \infty$.

Inspired by Eq.~\ref{eq-prob-dla} and keeping in mind the above considerations, here we define a new model for the 
growth of compact clusters, where no crossover in the radial growing probability is expected as $D_f = d$, which gives rise to a growing probability distribution $P_{RWG}(r,N) \propto r^{d-1}P(r,N)$.
The model is defined by assuming 1) that the cluster mass $N$ is related to the mean cluster radius $r_N$ by $N = (r_N/r_0)^{d}$ in $d$ spatial dimensions; and 2) that the radius of the cluster evolves as a random walker subject to a drift: at each updating step the radius varies of a quantity taken from a distribution with mean $v > 0$ (drift velocity) and variance $\sigma^2$.

Under these assumptions the radially averaged probability $P_{RWG}(r,N)dr$ that the $(N+1)$th cluster unit is deposited within a shell of width $dr$ at a distance $r$ from the center of mass of the cluster is
\begin{equation}
\label{eq-prob-rwcg}
P_\tn{RWG}(r,N) = \frac{r^k}{\mu^{(k)}_N}P(r,N),
\end{equation}
where $P(r,N)$ is given in Eq.~\ref{eq-prob-dla}, $\mu^{k}_N = \int_0^\infty dr r^k P(r,N)$ and we assume
$k = d-1$ as discussed above. The growth dynamics sets $\nu = 1/d$ and $\nu' = \nu/2 = 1/2d$.
Note that for $k<d-1$ the cluster density diverges when $r \to 0$, while for $k > d-1$ it does not decrease monotonically, and has a maximum at $r > 0$. 

\begin{figure}[t!!]
\begin{center}
\includegraphics*[scale=0.32]{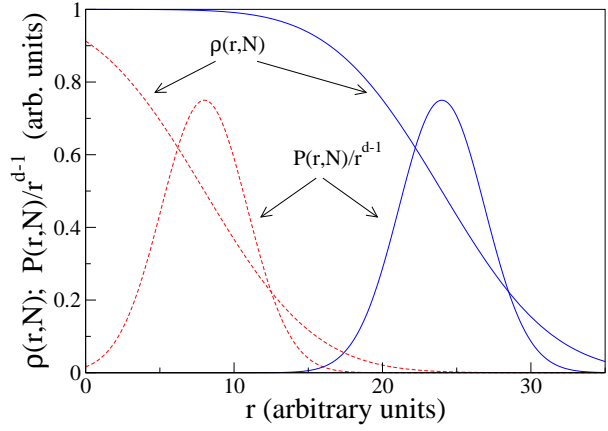}
\caption{\label{fig-schematic}
Model of cluster growth. The probability $P_\tn{RWG}(r,N)/r^{d-1}$ of building a cluster unit in a point at a distance $r$ from the center is Gaussian centered around the cluster radius $r_N=r_0 N^{1/d}$ with variance $\sigma_N^2 = \sigma^2 r_0 N^{1/d}/v$, Eq.~\ref{eq-prob-rwcg}. 
Here we plot in arbitrary units the cluster density $\rho(r,N)$ (Eq.~\ref{eq-prob-rwcg}) and the corresponding growing probability $P_\tn{RWG}(r,N)/r^{d-1}$ for $r_N = 8$ (dashed lines) and $r_N = 24$ (full lines) in $d=2$ spatial dimensions.
}
\end{center}
\end{figure}
We restrict our analysis to $d = 2$ dimensions, which is the more interesting case from a physical viewpoint, but an analytical treatment is also possible for $d = 3$.

According to Eqs.~\ref{eq-rho_dla} and \ref{eq-prob-rwcg} the density of a cluster of size $N = (r_N/r_0)^{d}$, is
\begin{eqnarray}
\label{eq-rho-1}
\rho(r,N)-\rho(r,0) &=& \frac{1}{S_d}\int_0^{N} \frac{1}{{r_n}^{d-1}}P(r,n)dn  \nonumber \\
&=& \frac{1}{S_d D_f v}\int_0^N\frac{n^{1/d-1}}{(2\pi)^{1/2}r^{d-1}_n \sigma_n}\exp\left[-\frac{(r-r_n)^2}{2\sigma_n^2}\right]dn,
\end{eqnarray}
and via the change of variables $y = n^{-1/2\alpha d}$,
\begin{eqnarray}
\label{eq-exact}
\rho(r,N)-\rho(r,0) &=& \sqrt{\frac{2}{\pi v r_0}}\frac{e^{\frac{v r}{\sigma^2}}}{2\pi\sigma} 
\int_{\sqrt{\frac{r_0}{r_N}} }^{\infty}y^{-2}\exp\left[-\frac{v}{2\sigma^2r_0}(r^2y^2-\frac{r_0^2}{y^2})\right]dy = \nonumber \\
&=& \frac{1}{4 \pi r_0 v}\left[2-\rm{Erfc}\left(\frac{r_N-r}{\sqrt{2} \sigma_N }\right)-e^{\frac{2 r r_N}{\sigma^2_N}} \rm{Erfc}\left(\frac{r_N+r}{\sqrt{2} \sigma_N }\right)\right].
\end{eqnarray}
The last term of Eq.~\ref{eq-exact} can be safely neglected in the case of large clusters, $r_N/\sigma_N \gg 1$~\footnote{A numerical evaluation of the usual L2 norm of $\rho(r,t)$ shows that an overestimate of the relative error made neglecting the last term in the case of real clusters (see Table~\ref{tavola}) is $10^{-2}$, which 
is obtained with $\lambda = \sigma^2/r_N = 0.6$ Km and $r_N=1$ Km.}, leading to the following approximation

\begin{equation}
\label{eq-approx}
\rho(r,N)\simeq\rho_\infty+\frac{\rho_\tn{max}-\rho_\infty}{2}\rm{Erfc}\left(\frac{r-r_N}{\sqrt{2} \sigma_N }\right),
\end{equation}
according to which the cluster density decays as a complementary error function.
Here $\rho_\infty = \rho(r,0)$ is the density before the growth of the cluster, or equivalently the density far away from the cluster center ($r \gg r_N$), and $\rho_\tn{max}-\rho_\infty = (2\pi r_0 v)^{-1}$. The maximum value of the density is achieved in the cluster center; for large clusters $r_N/\sigma_N = \sqrt{r_N v}/\sigma \gg 1$   and $\rho(0,N) \simeq \rho_\tn{max}$. A density profile similar to that of Eq.~\ref{eq-approx} is given by the solid-on-solid growth model in the high deposition limit~\cite{Weeks76}.

A schematic drawing explaining the growing dynamics of a cluster according to the proposed model is shown in Fig.~\ref{fig-schematic}. As a cluster grows $N$, $r_N$ and $\sigma_N$ increase;  $P_\tn{RWG}(r,N)$ shifts versus higher values of $r$ and becomes shorter and wider.

We do expect this model to be able to describe the growth of compact clusters, characterized by fluctuating and growing interface. This is the case of real cities, which grow as new buildings are constructed, usually in the suburbs, as we show in the following section.

\begin{figure}[t!!]
\begin{center}
\includegraphics*[scale=1]{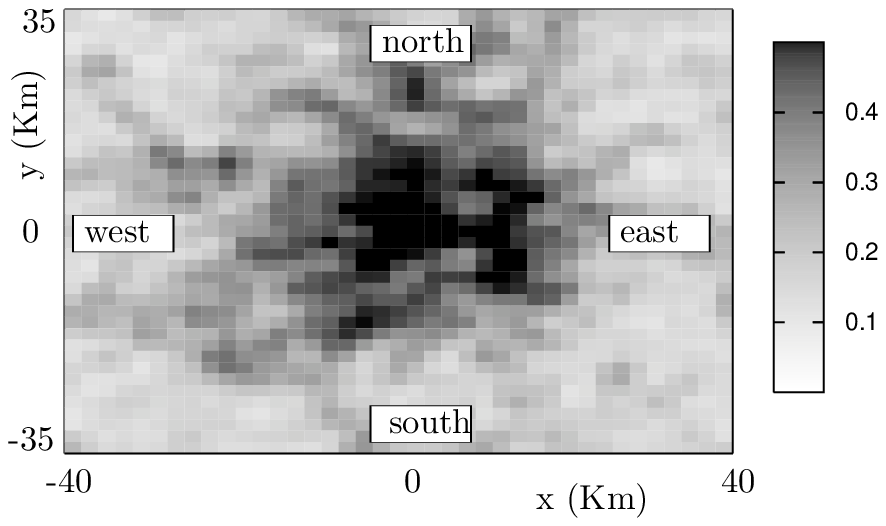}
\includegraphics*[scale=1]{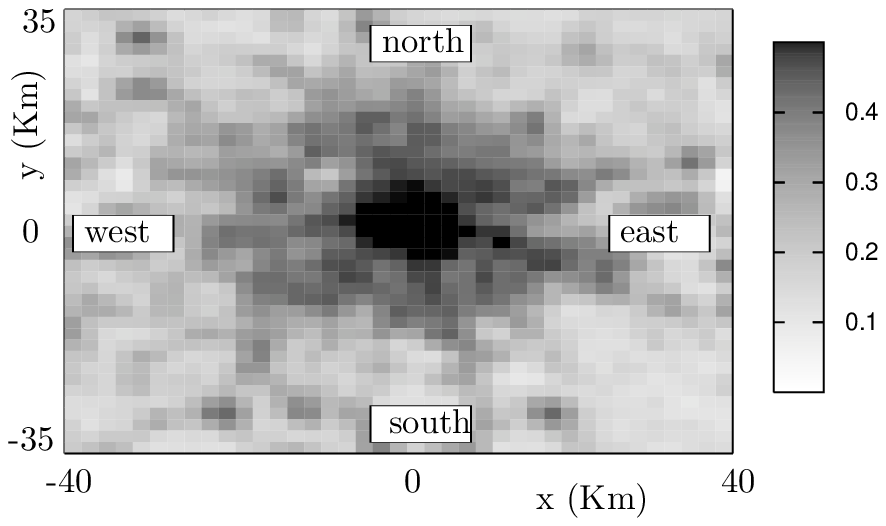}
\caption{\label{fig-rhos} The density of streets $\rho_s(x,y)$ of the city of Paris (upper panel) and of the city of London (lower panel). Each cell of this picture covers an area of size $1.9 \times 1.7$ Km$^2$ and indicates, according to the color code showed on the right, the percentage of the cell surface which corresponds to a street. The center of the city ($x = y = 0$) is located on the $\hat{\tn{I}}$le de la cit\'e for the case of Paris, and near Westminster for the case of London.
}
\end{center}
\end{figure}

\section{Empirical validation of the model}
We have validated the model by studying the radial dependence of the density of streets $\rho_s(r)$
of several European cities (here the subindex $s$ is used to indicate quantities related to streets), which has been obtained by analyzing digital maps~\cite{maporama} with a very high spatial resolution ($1$ pixel correspond to around $3.3$~m$^2$). The density of street $\rho_s^{A}$ of a given area $A$ is given by $\rho_s^{A} = n^A_s/n^A$, where $n^A$ is the total number of pixels which cover the are $A$, and $n^A_s$ is the number of pixels which cover the area $A$, and also lie on a street.
As far as we know this is the first empirical determination of the radial distribution of the density of streets, and no previous studies achieve such a high resolution in the determination of the morphology of a city.

\begin{figure}[t!!]
\begin{center}
\includegraphics*[scale=0.32]{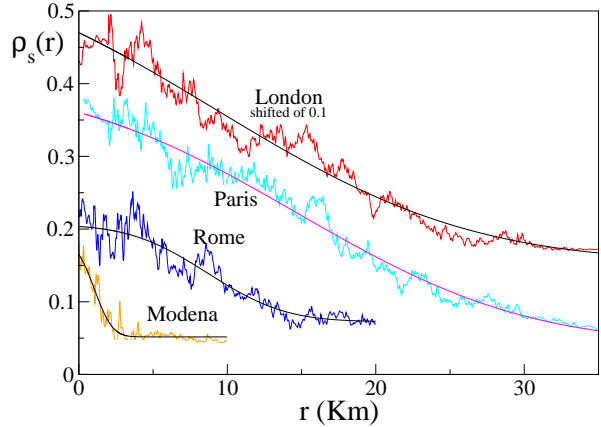}
\caption{\label{fig-tutte}
The density of streets as a function of the distance from the city center $\rho_s(r)$
for the cities of Modena, Rome, Paris and London. This last curve is shifted vertically of 0.1
for clarity. Plain lines are fit to Eq.~\ref{eq-exact}.
}
\end{center}
\end{figure}

As an example of the result of our digital image analysis we show in Fig.~\ref{fig-rhos} 
the density of streets $\rho_s(x,y)$ of the cities of Paris, France (upper panel) and London, U.K.
 (lower panel). For visualization purposes the images have a resolution $1.9 \times 1.7$ Km$^2$,
smaller than the actual resolution reached in our study. As expected the density of
streets reaches higher values in the city center ($x  \simeq y \simeq 0$), and smaller values in the
 suburbs. This is quantified in Fig.~\ref{fig-tutte} where we show the radial dependence of the 
density of streets $\rho_s(r)$ of several European cities (London (U.K.), Paris (France), Rome 
(Italy) and Modena (Italy)) which can be considered to a first approximation round. $\rho_s(r)$ 
appears to be always well fitted by the functional form of Eq.~\ref{eq-exact}. 

From the fitting parameters one can compute the radius of the city $r_N$, the variance $\sigma_N^2$ of the growing distribution or a characteristic growing length $\lambda = \sigma_N^2/r_N = \sigma^2/v$, and the values of the density of streets in the city center $\rho_{\rm max}$ and in the suburbs $\rho_\infty$. These values are reported in Table~\ref{tavola}. 
The parameter $r_N$ is a measure of the radius of a city, and therefore varies from city to city. The same is true for $\rho_{\rm max}$: Paris and London, which are characterized by the presence of large streets within the city center, have a higher value of $\rho_{\rm max}$ than Rome and Modena.
The parameter $\rho_0$ is the same for all considered cities, and therefore appears as a unifying property of the street network of the suburbs of all cities. Finally, the parameter $\lambda$ assumes similar values for the cities of Paris, Roma and Modena, and a higher value for the city of London. As $\lambda$ controls the variation of the variance of the distribution via $\sigma^2_N = \lambda r_N$ we conclude that London is less compact than other European cities.

\begin{table}[t!]
\begin{tabular*}{0.4\textwidth}
{|@{\hs{3mm}}c@{\hs{3mm}}|@{\hs{3mm}}c@{\hs{3mm}}|@{\hs{3mm}}c@{\hs{3mm}}|@{\hs{3mm}}c@{\hs{3mm}}|@{\hs{3mm}}c@{\hs{3mm}}|}
\cline{1-5}
City    & $r_N$ (Km)& $\lambda$ (Km)& $\rho_{\rm max}$ & $\rho_\infty$ \\ \cline{1-5}
Paris   & 17.7  & 0.31      & 0.36     & 0.06          \\ \cline{1-5}
London  & 14.5  & 0.54     & 0.36     & 0.05          \\ \cline{1-5}
Rome    & 8.7   & 0.31     & 0.20     & 0.07          \\ \cline{1-5}
Modena  & 1.2   & 0.34      & 0.16     & 0.05          \\ \cline{1-5}
\end{tabular*}
\caption{\label{tavola} Parameters characterizing the radial dependence of the density of streets of various European cities, obtained by making a fit of empirical data to Eq.~\ref{eq-exact}. All reported values have relative errors smaller than $5\%$.}
\end{table}

The possibility of describing the radial dependence of the density of streets of all considered cities with the same functional form (Eq.~\ref{eq-exact}) suggests the existence of a common mechanism underlying the growth of all cities,
which appears to be captured by the RWG cluster growth model.
The success of Eq.~\ref{eq-exact} in explaining the empirical data questions the common assumption of exponential decay of the population density~\cite{Clark51,Makse,Batty94}. In fact, (at least for $r \leq R$) it is reasonable to assume a proportionality between density of streets and population density (or density of urbanized area).

Our model reproduces the radially averaged properties of a city. In principle the radial dependency of Eq.~\ref{eq-exact} could be used in order to reproduce spatial correlations in urban settlements, as in~\cite{Makse}.

\section{Communication and congestion}
In many physical situations a cluster plays the role of a transportation infrastructure, as it is the backbone over which something moves. This is the case, for instance, of communication or transportation networks, both biological and artificial, which clusterize near `vital' centers. In these cases the topology of the cluster influences the dynamics of the transportation processes which occur over it~\cite{Trusina}. Particularly, large spatial gradients in the cluster area may lead to congestion and to a dramatic slowdown of the dynamics.
Here we discuss these influences in the case of clusters with a density given by the RWG model (Eq.~\ref{eq-exact}). We consider for definiteness the case of vehicular traffic over the street network of a city, but the results are more general.

Empirical evidences shows that there is a transition from free traffic flow
to congested traffic flow as the car density $\rho_{\rm car}$ increases. For instance highway traffic
becomes congested when $\rho_{\rm car} \geq \rho_{\rm car}^{\rm cong} \simeq 30$ vehicles/Km~\cite{Helbing01}. 
The knowledge of the density of streets allows for an estimate of the car density. Given $N$ cars in a region $A$ 
one can estimate $\rho_{\rm car} = \beta N/S_A$ where $S_A = \int_A \rho_s(x,y) dx dy$ is the area occupied by the streets in the region $A$, and $\beta$ a coefficient of proportionality.

Let's consider what happens when $N$ cars, initially located in a circular annulus of radius $r$ and width $\Delta$ around the city, move outward. 
\begin{figure}[!!t]
\begin{center}
\includegraphics*[scale=0.32]{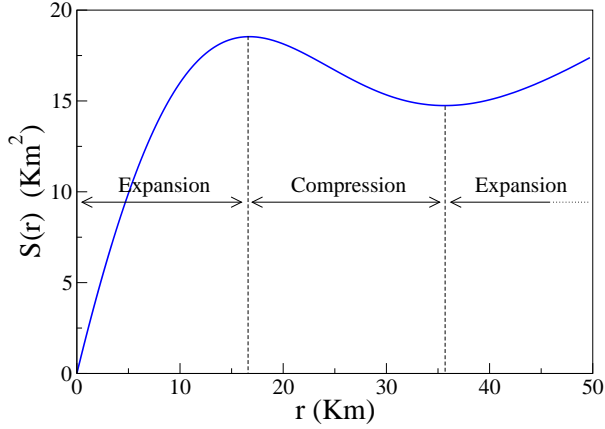}
\caption{\label{fig-sr}
Area occupied by the streets in a circular annulus of width $\Delta = 1$ Km centered on the city center, $S(r) = 2 \pi r \Delta \rho_s(r)$, as a function of the distance from the city center $r$. The curve refers to the city of London, but similar curves are found for the cities of Paris, Rome and Modena. The non-monotonic behavior of $S(r)$ influences vehicular traffic:  cars moving outward are compressed in the region $17 \lesssim r \lesssim 36$ Km where $S(r)$ decreases.
}
\end{center}
\end{figure}
As their mean distance from the city center $r$ increases the car density varies as $\rho_{car}(r) = \beta N/S(r)$, where 
\begin{equation}
S(r) = 2 \pi \int_r^{r+\Delta} \!\!\!\!\!\!\!\! r'\rho_s(r')dr'  
2 \pi r \Delta  \simeq\left[\rho_\infty+\frac{\rho_\tn{max}-\rho_\infty}{2}\rm{Erfc}\left(\frac{r-r_N}{\sqrt{2} \sigma_N }\right)\right]
\end{equation}
is the area occupied by the streets in the circular annulus we are considering. The car density is higher where $S(r)$ is smaller, and increases in the regions in which $\partial S/\partial r < 0$. These regions are those which most probably act as traffic bottlenecks.

For the city of London, Figure~\ref{fig-sr} shows that $S(r)$ increases for $r \lesssim 17$ Km and for $r \gtrsim 36$ Km, when the density of streets varies very slowly. In the range $17 \lesssim r \lesssim 36$ Km of fast variation of the density of streets $S(r)$ decreases. It follows that in this range of distances from the city center, i.e. at the city boundary, traffic is maximum. This is therefore the area where new transportation infrastructures and strategies for traffic optimization should be concentrated. The non-monotonic behavior of $S(r)$
also characterize Paris, Rome and Modena.

\section{Conclusions}
In this paper we have introduced and studied a new model of cluster growth. This model is inspired by numerical results relative to the DLA and to the Eden model, but it takes into account the fact that the probability of building a cluster unit in a given point depends on the local density. In this model the probability of building a cluster unit in a point at a distance $r$ from the cluster center is a Gaussian centered around the radius of the cluster, with variance proportional to the radius, modulated by the local density $\rho(r)$. An explicit solution of the cluster properties in $d=2$ dimensions, which is not available for other cluster growth models, shows that the cluster density decays as complementary error function. We have validated the model via a high resolution study of the density of streets of several European cities, and we have discussed the relation between the cluster topology and the dynamics over the cluster.

\ack
Work supported by EU Network Number MRTN-CT-2003-504712, MIUR-PRIN 2002, MIUR-FIRB 2002, CrdC-AMRA, INFM-PCI.

\end{document}